\documentclass[epj]{svjour}
\usepackage{graphicx}
\usepackage{pslatex}
\usepackage{color}
\usepackage{xspace}

\graphicspath{{figs/}}

\def\FastJet{{\sc FastJet}\xspace}
\def\mur{\ensuremath{\mu_r}\xspace}
\def\muf{\ensuremath{\mu_f}\xspace}
\def\mqPole{\ensuremath{m_q^{\mbox{\scriptsize pole}}}\xspace}
\def\mtPole{\ensuremath{m_t^{\mbox{\scriptsize pole}}}\xspace}
\def\mqMS{\ensuremath{\overline{m}_q(\mur)}\xspace}
\def\GeV{\textnormal{GeV}}
\def\as{\ensuremath{\alpha_s}\xspace}
\def\ttbar{\ensuremath{t\bar t}\xspace}
\def\ttbaronejet{\ensuremath{t\bar t + 1\textnormal{\normalsize -jet}}\xspace}
\def\sigmattj{\ensuremath{\sigma_{t\bar t + \textnormal{\scriptsize
        1-jet}}}\xspace}
\def\sttj{\ensuremath{s_{t\bar t j}}\xspace}
\def\n3{\ensuremath{{\cal R}}\xspace}
\def\rhos{\ensuremath{\rho_s}\xspace}
\def\pt{\ensuremath{p_T}\xspace}
\def\MSbar{\ensuremath{\overline{\mbox{MS}}}\xspace}
\def\Tab#1{Tab.~\ref{#1}}
\def\Fig#1{Fig.~\ref{#1}}
\def\Figs#1{Figs.~\ref{#1}}
\def\Eq#1{Eq.~(\ref{#1})}
\def\Ref#1{Ref.~\cite{#1}}
\def\Refs#1{Refs.~\cite{#1}}

\def\makeheadbox{{%
\hbox to0pt{\vbox{\baselineskip=10dd\hrule\hbox
to\hsize{\vrule\kern3pt\vbox{\kern3pt
\hbox{\sffamily DESY 13-014, HU-EP-12/30, IFIC/13-08, LPN 13-011, SFB/CPP-13-06}
\kern3pt}\hfil\kern3pt\vrule}\hrule}%
\hss}}}

\begin{document}
\title{A new observable to measure the top-quark mass at hadron colliders}
\author{S. Alioli\inst{1} \and  P.~Fernandez\inst{2} 
  \and J. Fuster\inst{2} \and A. Irles\inst{2} 
  \and S. Moch\inst{3,4} \and P. Uwer\inst{5} \and M. Vos\inst{2}
}      

\institute{
  LBNL \& UC Berkeley, 1 Cyclotron Road, Berkeley, CA 94720, USA 
  \and 
  IFIC, Universitat de Val\`encia and CSIC, Catedr\'atico Jose
  Beltr\'an 2, E-46980 Paterna, Spain 
  \and 
  II. Inst. f\"ur Theoretische Physik, Universit\"at Hamburg, 
  Luruper Chaussee 149, D-22761 Hamburg, Germany 
  \and 
  DESY, Platanenallee 6, D-15738 Zeuthen, Germany 
  \and 
  Humboldt-Universit\"at zu Berlin, Newtonstrasse 15, D-12489
  Berlin, Germany
}
%
%
\abstract{ A new method to measure the top-quark mass in high
  energetic hadron collisions is presented. We use theoretical
  predictions calculated at next-to-leading order accuracy in quantum
  chromodynamics to study the (normalized) differential distribution
  of the \ttbaronejet cross section with respect to its invariant mass
  $\sqrt{\sttj}$. The sensitivity of the method to the top-quark mass
  together with the impact of various theoretical and experimental
  uncertainties has been investigated and quantified.  The new method
  allows for a complementary measurement of the top-quark mass
  parameter and has a high potential to become competitive in
  precision with respect to established approaches. Furthermore we
  emphasize that in the proposed method the mass parameter is uniquely
  defined through one-loop renormalization.  }

\maketitle

\section{Introduction}
\label{intro}
With a mass of $173.2\pm 0.9$ GeV the top quark is the heaviest
elementary fermion discovered so
far~\cite{Aaltonen:2012ra,Abazov:2011pta,Beringer:1900zz}.  In the
Standard Model (SM) the large mass constrains the top-quark lifetime
to become extremely short, inhibiting top-quark bound states to be
formed. As an important consequence top quarks offer the unique
possibility to study the properties of a quasi-free quark. Owing to
its parity violating decay it is, for example, possible to analyze the
top-quark polarization in difference to the lighter quarks where the
spin information is typically diluted through the hadronization. In
the SM the top-quark Yukawa coupling is very close to one.  With the
strongest coupling to the Higgs boson the top quark represents an
ideal laboratory for detailed tests of the Higgs mechanism. Beyond the
SM the top quark plays an important role in scenarios aiming to give
an alternative explanation of spontaneous electroweak symmetry
breaking (EWSB). Top-quark physics is thus a sensitive probe for
precision tests of the SM but also for new physics searches. The Large
Hadron Collider (LHC) will allow measurements of the top-quark
properties with unmatched accuracy. In the SM the top-quark couplings
are completely predicted through the gauge structure. The only free
parameters in the top-quark sector are the matrix elements of the
Cabbibo-Kobayashi-Maskawa (CKM) mixing matrix and the top-quark mass.
The CKM matrix elements are highly constrained through indirect
measurements. The LHC experiments will complement this picture through
direct measurements of the CKM matrix elements in single top-quark
production.  Precise top-quark mass measurements are of high physics
relevance. First, on its own right as the top-quark mass represents a
fundamental parameter of the SM. Second, because its
comparison with the mass of the recently observed resonance
\cite{atlashiggs:2012gk,cmshiggs:2012gu} ---assuming that the
resonance is the long-sought Higgs boson---can be used to test the
validity of the SM \cite{Heinemeyer:2006px} since the $W$
boson mass, the top-quark mass and the Higgs boson mass
are related in the SM. Very recently also the impact of the top-quark
mass on the stability of the
electroweak vacuum has been revisited~\cite{Degrassi:2012ry,Alekhin:2012py}.

Quarks, in contrast to leptons, are colored objects which interact
strongly in addition to the usual electroweak interactions. As a
consequence quarks are not observed as free particles. They are
confined into colorless hadrons. The natural way to measure quark
masses is thus to treat them similarly to other couplings of
the underlying theory: {\it the couplings are measured through their
  influence on hadronic observables}. More precisely, theoretical
predictions are compared with measurements and the couplings are then
obtained through a fit.  For two examples where this idea has been
applied in practice we refer to
\Refs{Bilenky:1998nk,Langenfeld:2009wd}.  Since the parameters of a
model are in general not observables by themselves their precise
values depend on the renormalization scheme used to define them.
Qualitatively renormalization controls which part of quantum
corrections are already accounted for through the renormalized
couplings measured in the experiments. As a consequence one needs at
least a next-to-leading order (NLO) calculation---where quantum
corrections appear for the first time in the theoretical
predictions---to define unambiguously the renormalization scheme.  In
leading order (LO) different schemes cannot be distinguished and are
formally equivalent in perturbation theory.  In the case of heavy
quark masses the most commonly used definitions are the pole mass
\mqPole and the running mass \mqMS. The former is defined as the pole
of the renormalized quark propagator while the latter is defined
through modified minimal subtraction (\MSbar). In the \MSbar scheme
the renormalization constants are chosen such that only the
ultraviolet divergences together with the constant $-\gamma_E +
\ln(4\pi)$ are removed through the renormalization program.  (The
constant $\gamma_E=0.577215\ldots$ denotes the Euler-Mascheroni
constant.)  Similar to the coupling constant of the strong
interaction, the running mass \mqMS depends on the renormalization
scale \mur.  Physics should be independent of the mass definition
used. However at any fixed order in perturbation theory a
residual dependence on the mass definition as well as on the arbitrary
renormalization scale \mur remains.  In specific cases, the \mur
dependence of the \MSbar mass can be used to absorb certain
logarithmically enhanced corrections and resum them to all orders
within perturbation theory.  This may lead to an improved behavior of
the perturbative predictions (see for example
\Ref{Langenfeld:2009wd}). In top-quark physics the large top-quark
mass naturally sets a large scale resulting in a small value of the
QCD coupling \as evaluated at $\mur = \mtPole$ and we believe that the
aforementioned scheme dependence is in general less important than for
the lighter quarks.  Nevertheless, if precise determinations
of the top-quark mass are pursued, it is of major relevance
to understand the quantitative extent of this qualitative
argument. A discussion of different methods and approaches
used to measure the top-quark mass will contribute in
clarifying this statement.

Direct determinations of the top-quark mass have been performed at the
Tevatron and LHC colliders. The top-quark mass is presently inferred
by the kinematical reconstruction of the invariant mass of its decay
products with techniques such as the matrix element or the template
method (see e.g.,~\Ref{Aaltonen:2012ra} and references therein) or by its
relation to the top-quark pair production cross
section~\cite{Langenfeld:2009wd}.  The top-quark mass derived from the
kinematical reconstruction does not correspond to a well-defined
renormalization scheme leading to a theoretical  uncertainty in its
interpretation. Nevertheless it is usually interpreted as the
top-quark pole mass \mtPole. The present results achieved using the
kinematical reconstruction reach a higher precision ($m_t=173.2\pm
0.9$~GeV~\cite{Aaltonen:2012ra}) than those which are extracted from the
cross section measurements (e.g. $\mtPole=173.3 \pm
2.8$~GeV~\cite{Alekhin:2012py} using \Ref{Aliev:2010zk}).  The large
experimental uncertainty of the mass determinations based on cross
section measurements is a consequence of the sensitivity of the
cross section on the top-quark mass.  However we emphasize that in this
measurement the renormalization scheme is unambiguously defined in
difference to the determination based on the kinematical
reconstruction. In particular using the cross section measurements a
direct extraction of the running mass is possible. This was for the
first time done in \Ref{Langenfeld:2009wd} and is currently repeated
by the CMS and ATLAS collaborations.  The experimental
accuracy of the method is limited by the weak sensitivity of the cross
section on the top-quark mass.  Assuming an
experimental accuracy of 5\% on the cross section measurements a
determination of the mass accurate to 1\% could be envisaged.

Given the physical relevance of top-quark mass measurements  it is
important to employ alternative methods with similar or even better
accuracy than the methods mentioned above. With the current accuracy
of the mass measurements being better than one per cent this is a
highly non-trivial task.  In this work we advocate a new method to
measure the top-quark mass in high energetic hadron collisions at the
LHC. The mass dependence of the production of top-quark pairs in
association with an additional jet is exploited.  This process is
sensitive to the top-quark mass since gluon radiation depends on the
top-quark mass through threshold and cone effects. More
  precisely, we study the normalized \ttbaronejet cross section
differential in the invariant mass of the final state jets.  In
section \ref{sec:2} a detailed description of the observable is given
together with a short discussion of the available theoretical
predictions including an analysis of the theoretical uncertainties.
Section \ref{sec:3} reports on a generic study of the observable at
the particle level. In particular the major uncertainties of the
method are investigated.  We finally present our conclusions in
section \ref{sec:conclusions}.

\section{Top-quark pair production in association with a hard jet at
  NLO accuracy in QCD}
\label{sec:2}

The production process for top-quark pairs in association with a hard
jet occurs with a large rate at the LHC.  Together with the high
statistics data samples based on the integrated luminosity of the
runs at 7 and 8 TeV, the cross section for \ttbaronejet
will become a precision measurement.  For the experiments, this
provides the opportunity for detailed studies of differential
distributions as well as an accurate determination of the top-quark
mass.  On the theory side, it necessitates the computation of radiative
corrections to sufficient accuracy.

The NLO QCD corrections for $\ttbaronejet + X$ have been presented in
\Refs{Dittmaier:2007wz,Dittmaier:2008uj}.  The results share all the
attractive features of theoretical predictions including radiative
corrections at higher orders.  First of all, they lead to a
significant reduction of the renormalization/factorization scale
uncertainty compared to the LO predictions.  Secondly, they display
the apparent convergence of the perturbative expansion.  It is a
remarkable result of \Refs{Dittmaier:2007wz,Dittmaier:2008uj} that the
NLO corrections are numerically small.  The one-loop corrections
change the $\ttbaronejet + X$ cross section by less than 15\%.  To
estimate the effect of uncalculated higher order corrections the
standard procedure is adopted in
\Refs{Dittmaier:2007wz,Dittmaier:2008uj}. The factorization scale \muf
and the renormalization scale \mur are set equal and varied up and
down by a factor of two. As central scale the top-quark mass is used
with the mass renormalized in the pole mass scheme.  Thus, the cross
section for $\ttbaronejet + X$ is theoretically very well under
control and well suited for precision measurements. We also note that
the scale dependence is rather flat around the central value.
Independent variations of the renormalization and
factorization scales have been considered, for example, in
\Ref{Langenfeld:2009wd} for the NLO and NNLO $t \bar t + X$
production and in \Ref{Alioli:2011as} for the $\ttbaronejet + X$
process at NLO. The outcome of these studies is that independent
scale variations---excluding extreme combinations resulting in an
overall factor greater than two or smaller than one half---do not
significatively change the results obtained considering simultaneous
variations.

In \Refs{Dittmaier:2007wz,Dittmaier:2008uj} results for both the
Tevatron and the LHC collider were presented, in the latter case for
the nominal design energy of 14 TeV.  Since the LHC was operating at 7
TeV in the first run period we have updated the results of
\Refs{Dittmaier:2007wz,Dittmaier:2008uj} to 7 TeV.  We have not yet
updated the results to account for the 8 TeV center-of-mass energy
used in the 2012 LHC run because we do not expect a substantial change
of the results presented in this work.  The LHC results for 7 TeV are
given in \Tab{tab:LOandNLOxsections}. In difference to
\Refs{Dittmaier:2007wz,Dittmaier:2008uj} a different jet algorithm has
been applied.  While \Refs{Dittmaier:2007wz,Dittmaier:2008uj} used the
kt-algorithm \`a la Ellis and Soper \cite{Ellis:1993tq} we have
applied the anti-kt algorithm \cite{Cacciari:2008gp} as implemented in
\FastJet \cite{Cacciari:2011ma}.  The $R$ value is set to 0.4---a
value which is also used in the experimental analysis. The
recombination scheme used is the $E-$scheme.  To render the cross
section infrared finite we demand a minimum \pt of 50 GeV for the
light jet.
\begin{table}[h]
\begin{center}
\caption{The $\ttbaronejet + X$ cross section using
  LO and NLO calculations \cite{Dittmaier:2007wz,Dittmaier:2008uj} for
  proton-proton collisions at 7 TeV and for different \mtPole values.
  Jets are defined using the anti-kt algorithm \cite{Cacciari:2008gp}
  with R=0.4 as implemented in the \FastJet package
  \cite{Cacciari:2011ma}. The additional jet is required to have
  $\pt>50$~GeV and $|\eta|<2.5$. The uncertainty due to the limited
  statistics of the numerical calculation is indicated in parenthesis
  affecting the last digit. The scale uncertainty is also shown for
  some top-quark mass values. The CTEQ6.6 \cite{Nadolsky:2008zw}
  (CT09MC1 \cite{Lai:2009ne}) PDF set has been used to obtain the NLO
  (LO) results.
  \label{tab:LOandNLOxsections}}
\renewcommand{\arraystretch}{1.3}
\begin{tabular}{l|l|l}
\hline
 & \multicolumn{ 2}{c}{\sigmattj [pb]}\\
 & \multicolumn{ 2}{c}{$\pt(jet)>50 \,\mbox{GeV}, \,|\eta(jet)|<2.5$    }\\
\hline
\mtPole [GeV] & LO & NLO\\
\hline
$160$ & $66.727(5)$ & $60.04(8)$ \\
$165$ & $57.615(4)$ & $52.25(9)$ \\
\textbf{$170$} & \textbf{$49.910(3)^{+30}_{-17}$} & \textbf{$45.45(6)^{+1}_{-6}$
} \\ 
\textbf{$172.5$} & \textbf{$46.508(3)^{+28}_{-15}$} & \textbf{$42.37(6)^{+1}_{-6
}$} \\ 
$175$ & $45.372(3)$ & $39.46(6)$ \\
$180$ & $37.800(2)$ & $34.73(5)$ \\
\hline
\end{tabular}
\end{center}
\end{table}
For the parton distribution functions (PDFs) we have used
CTEQ6.6~\cite{Nadolsky:2008zw} as the default PDF set. Note that this
set does not include LO parton distribution functions. For the LO PDFs
we used the CT09MC1 \cite{Lai:2009ne} set as recommended by the CTEQ
collaboration\footnote{Private communication with Pavel Nadolsky}.  In
the LO predictions we use for the QCD coupling constant $\alpha_s$
the value provided by the CT09MC1 set. For the LO contributions entering
the NLO predictions we consistently use the NLO PDF set together with the
corresponding $\alpha_s$ value.  Independent on the top-quark mass, we find
negative corrections of about 10\%. As mentioned above the results
shown in \Tab{tab:LOandNLOxsections} cannot be directly compared with the
results given in \Refs{Dittmaier:2007wz,Dittmaier:2008uj} owing to the
different collider energy, different PDF set and the different jet
algorithm.
The results of the scale variation around
$\mu=\mur=\muf=\mtPole$ are shown as sub- and superscript and the
top-quark mass is renormalized in the pole mass scheme. The subscript
denotes the shift of the cross section for $\mu=2\mtPole$ and,
likewise, the superscript for $\mu=\mtPole/2$.  In comparison to LO
the scale uncertainty is significantly reduced in agreement
with \Refs{Dittmaier:2007wz,Dittmaier:2008uj}.

\Tab{tab:LOandNLOxsections} also illustrates the mass dependence. Similar to
what has been observed in inclusive top-quark pair production we find
\begin{equation}
  {\Delta \sigmattj \over \sigmattj} \approx  -5  {\Delta \mtPole\over
    \mtPole}\, ,
  \label{eq:ttjxsectionsensitivity}
\end{equation}
which indicates that a measurement of the $\sigmattj$ cross section 
accurate to 5\% would imply an uncertainty in the mass of 1\%.
In section \ref{sec:Observable} we will show how this sensitivity can
be improved by considering differential distributions.   

Finally, the uncertainties originating from the PDFs 
have been studied by comparing the results obtained with the
PDF sets CTEQ6.6 and MSTW2008nlo90cl~\cite{Martin:2009iq}.
For a top-quark mass of $\mtPole = 170$~GeV we find, for example,
\begin{equation}
  \sigmattj^{\mbox{\scriptsize NLO, MSTW08}} = 49.21 ~ {\rm pb}
  \, .
\end{equation}
We observe a sizeable difference of about 10 per cent between the two
sets, see \Ref{Alekhin:2012ig} for a discussion of benchmark cross
sections at the LHC and PDF uncertainties.
However, as we shall see later, the PDF uncertainties will largely
cancel in normalized distributions. 
\begin{table}[h]
\begin{center}
\caption{The $\ttbaronejet + X$ cross section for proton-proton
  collisions at 7 TeV obtained with the POWHEG-BOX
  \cite{Alioli:2010xd,Frixione:2007nw,Alioli:2011as} for $\mtPole=170$ GeV. 
  The setup is the same as in \Tab{tab:LOandNLOxsections}.}
\label{tab:PSxsections}
\renewcommand{\arraystretch}{1.3}
\begin{tabular}{l|l|l|}
  \hline
  &
  \multicolumn{2}{c}{ \sigmattj [pb]} \\[0.5ex]
 \hline
  \multicolumn{ 1}{c|}{$t\bar{t}$} & without additional PS & 
  \multicolumn{ 1}{c}{50.42(6)} \\ \cline{ 2- 3}
  \multicolumn{ 1}{c|}{NLO}  & PS by Pythia8 & \multicolumn{ 1}{c}{45.61(8)} \\ 
 \hline
  \multicolumn{ 1}{c|}{$t\bar{t}+\mbox{1-jet}$} & without additional
  PS & 
  \multicolumn{ 1}{c}{48.8(2)} \\ \cline{ 2- 3}
  \multicolumn{ 1}{c|}{NLO} & PS by Pythia8 & \multicolumn{ 1}{c}{45.1(1)} \\ 
  \hline
\end{tabular}
\end{center}
\end{table}
Very recently predictions for \ttbaronejet production matched with
parton shower predictions (PS) have been provided
\cite{Alioli:2011as,Kardos:2011qa}.  We have used the results from
\Refs{Alioli:2011as,Frixione:2007nw} to investigate the effect of the
parton shower and allowing for a more realistic study closer to what
will be done in the experimental analysis.  In \Tab{tab:PSxsections}
results for $\mtPole=170$ GeV are shown. For completeness, we also
include the inclusive top-quark pair production simulated with POWHEG.
In the inclusive prediction the additional jet is included in NLO
accuracy through the real corrections and one could argue that
together with the parton shower description this approach should give
already a reasonable description of \ttbaronejet production.  For both
cases, $t\bar t$@NLO + POWHEG and $\ttbaronejet$@NLO + POWHEG, we
observe that the matching with the Pythia8 parton shower
\cite{Sjostrand:2007gs} leads to a reduction of the predicted cross
sections, partially due to the effect of the selection cuts.  In case
of \ttbaronejet production the results obtained within the POWHEG
framework are in perfect agreement with the fixed order results shown
in \Tab{tab:LOandNLOxsections} once the Pythia8 parton shower is
included.

\begin{figure*}
\centering
\includegraphics[width=16cm]{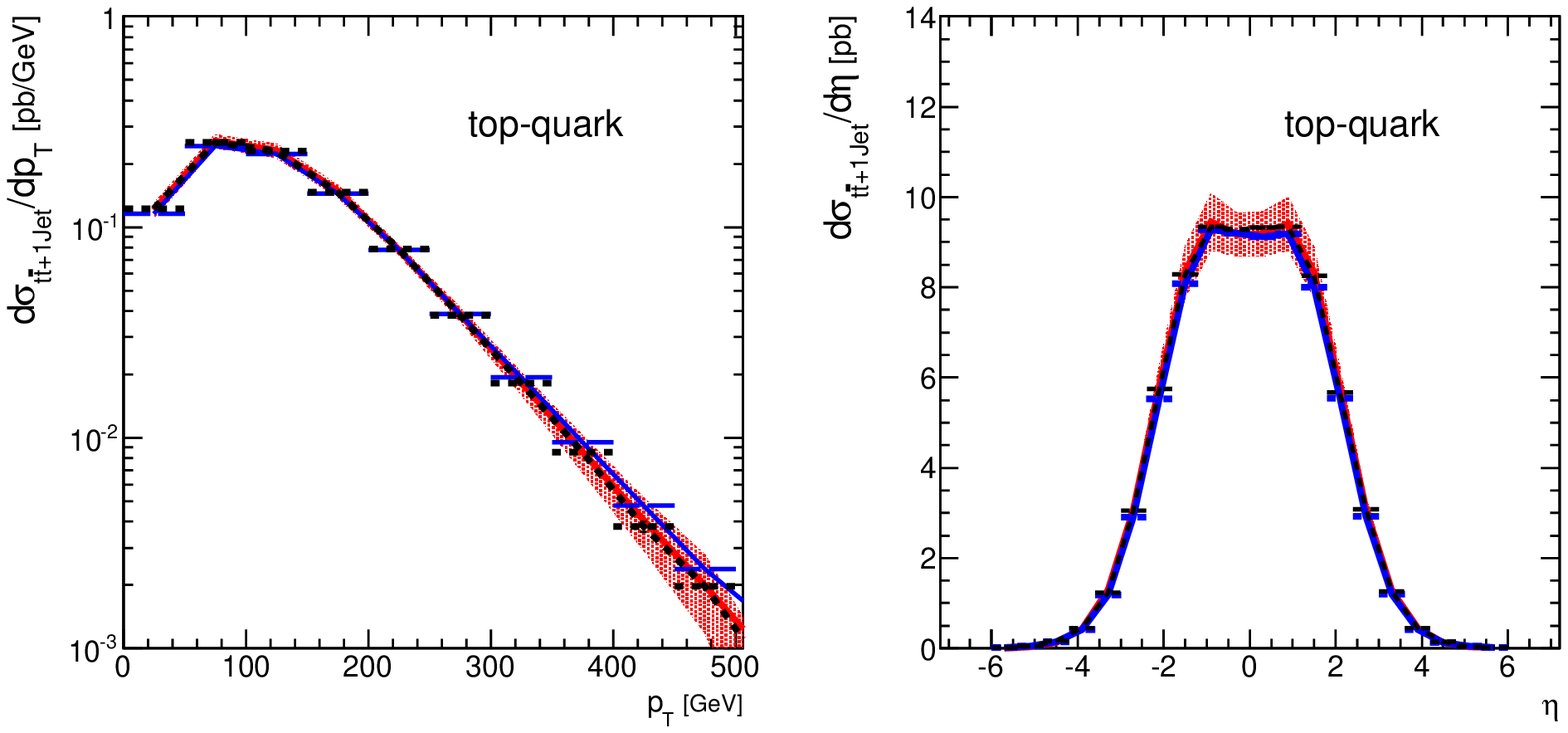}
\includegraphics[width=16cm]{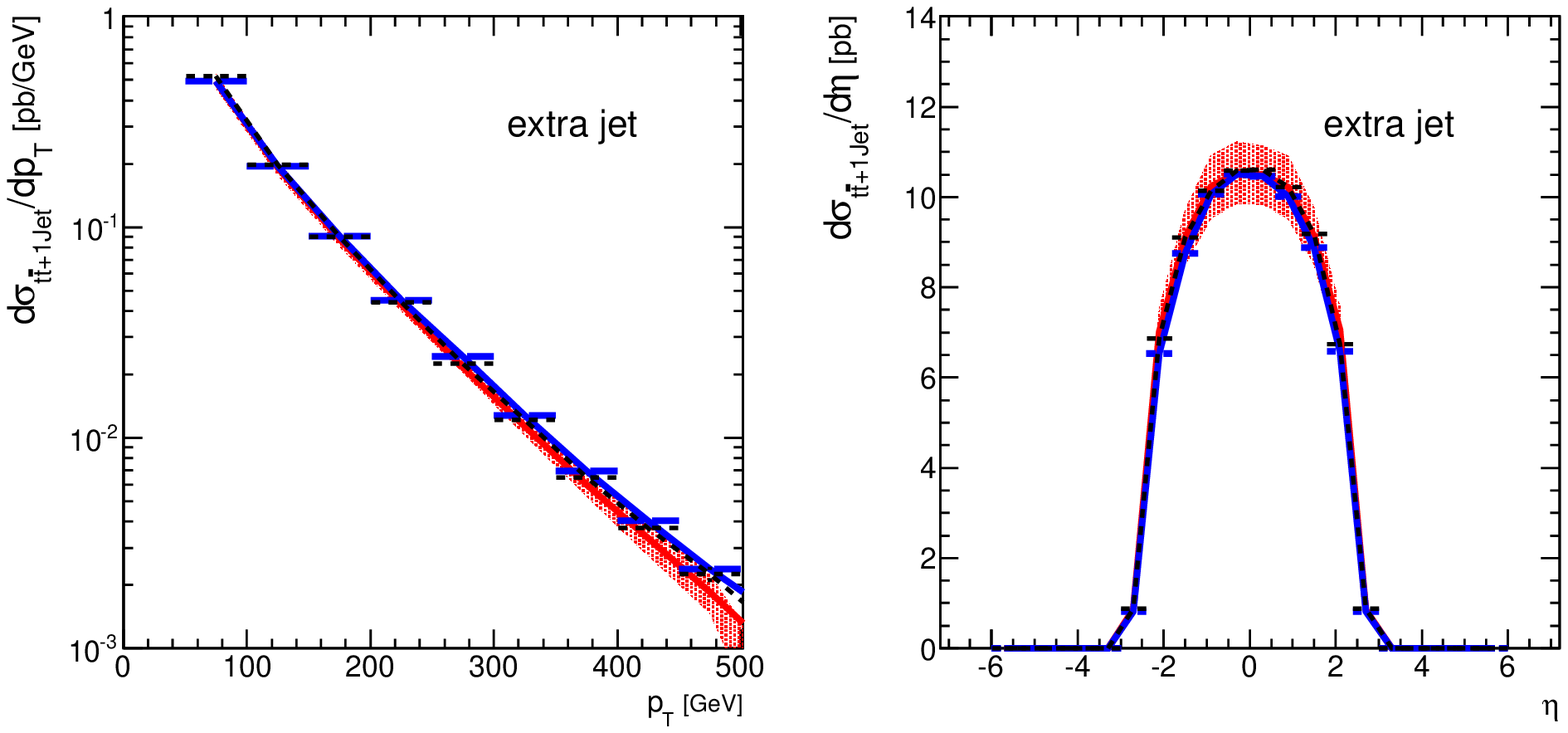}
\includegraphics[width=16cm]{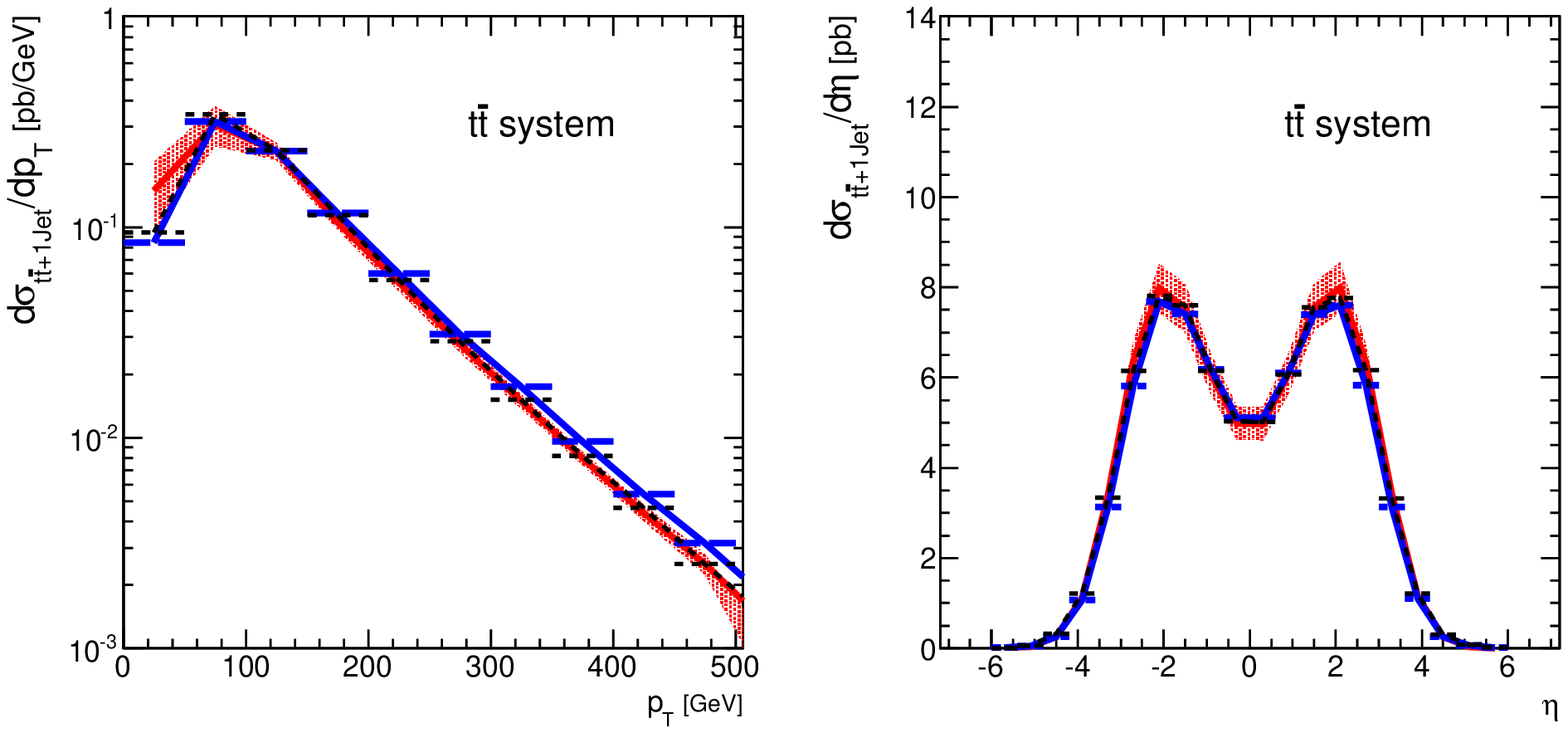}
\vspace*{0.1cm}

\caption{
  Comparison of different theoretical approaches to
  describe the \ttbaronejet production applied to various \pt and $\eta$
  distributions. The red band corresponds to the \ttbaronejet NLO at
  fixed order including the scale uncertainty. The continuous-blue
  and dotted-black line show the results obtained using POWHEG with
  \ttbar and \ttbaronejet NLO calculations, respectively, and matched
  with the Pythia8 parton shower.}
\label{fig:pt_pythia}
\end{figure*}

To further investigate the reliability of the theoretical predictions
we have also analyzed differential distributions obtained within
different approaches. This is shown in \Fig{fig:pt_pythia}. In detail
we compare distributions obtained within the POWHEG framework with those
results obtained in the fixed order parton level calculation.
The POWHEG results were obtained in combination with the
Pythia~8.150\, parton shower.  In general the agreement between
different approaches is very good.  Only at large transverse momentum
we observe minor differences.  While the fixed-order parton level
results agree well with \ttbaronejet{}@NLO + POWHEG, the results
obtained with $t\bar t$@NLO + POWHEG are slightly larger.  Since the
hard emission in $t\bar t$@NLO + POWHEG is only treated in LO, this
could be a consequence of the missing double real emission processes.
In $t\bar t$@NLO + POWHEG any additional jet activity beyond the first
emission is entirely due to the parton shower which prefers soft and
collinear emissions.  In the fixed-order \ttbaronejet calculation and
also in the corresponding POWHEG implementation a second
hard parton---which may be clustered in a second jet---is
included already in the hard matrix elements. The effect of this hard
emission can easily explain the reduced average transverse momentum at
large \pt\footnote{We observe that in this distribution also
    the argument of $\alpha_s$ plays a role in determining the
    spectrum: if the second jet is generated by the shower, typically
    there would be an $\alpha_s(\pt)$ factor associated to that
    emission. Instead, when computed with the exact matrix element,
    our choice of $\mur=m_t$ will results in a larger $\alpha_s$ value
    and thus an harder spectrum, for $\pt > m_t$.}. In case of the
transverse momentum distribution of the $t\bar t$ system we observe
differences at low $p^{t\bar t}_T$ between the fixed-order calculation
and the two POWHEG implementations.  This is not unexpected, because
it is precisely the region where the parton shower approach resums
logarithmically enhanced corrections to all orders.  Due to
the additional cut which we have imposed on the light jet this does
not occur in the transverse momentum distribution of the light jet.

The various studies presented above show, that the
theoretical description of the \ttbaronejet process
at NLO accuracy in QCD is well under control.

\subsection{Top-quark mass measurements with \ttbaronejet events}
\label{sec:Observable}
As shown in \Eq{eq:ttjxsectionsensitivity}, the mass sensitivity of
the \ttbaronejet cross section \sigmattj is very similar to the
inclusive $t\bar t$ cross section.  Therefore, most likely, a
measurement of the `inclusive' \ttbaronejet cross section would not
lead to any significant improvement compared to the mass measurements
already performed in inclusive top-quark pair production.  Instead, we
would like to propose an alternative approach.

\begin{figure}
\includegraphics[width=\columnwidth]{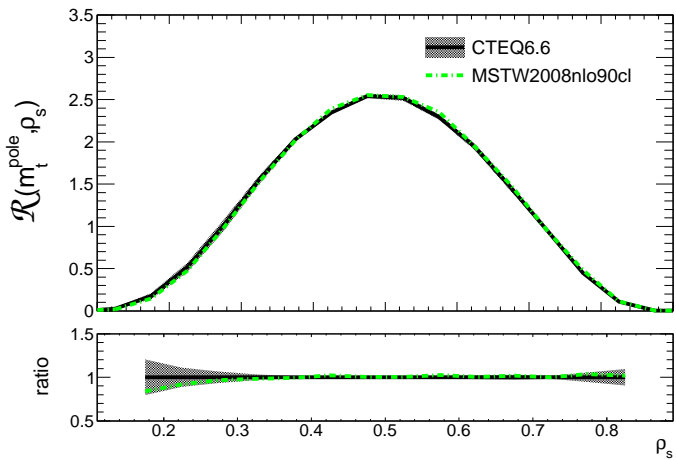}
\caption{Predictions for \n3 at NLO accuracy using two different PDF
  sets (CTEQ6.6, MSTW2008nlo) for $\mtPole=170$~\GeV. For CTEQ6.6
  the uncertainty due to scale variation is shown as band. The ratio
  between both predictions is shown together with the scale
  uncertainty.}
\label{fig:PDFdependence}
\end{figure}

\begin{figure}
  \includegraphics[width=\columnwidth]{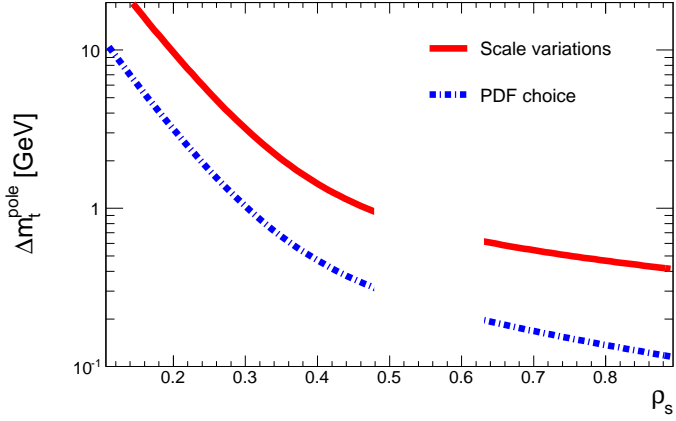}
  \caption{Expected impact of scale (magenta line) and PDF (blue
    dashed line) uncertainties on the measured top-quark mass value.
    The region where \n3 is essentially insensitive to the top-quark
    mass is not shown.
    \label{fig:Uncertainties}}
\end{figure}

Since inclusive cross sections are in general difficult to measure, we
propose to study normalized differential distributions. Evidently
distributions contain more information and may be more sensitive to
the mass parameter. Furthermore, due to the normalization many
experimental and theoretical uncertainties cancel between numerator
and denominator.  In order to enhance the mass sensitivity of
\ttbaronejet events we need to focus on kinematical configurations
where an enhanced sensitivity can be expected.  A natural observable
to look at is the (normalized) differential distribution of the
\ttbaronejet cross section with respect to the invariant mass squared
\sttj of the final state. More precisely we study the dimensionless
distribution
\begin{equation}
\label{eq:n3Definition}
\n3(\mtPole,\rhos)= 
\frac{1}{\sigmattj} 
\frac{d\sigmattj}{d\rhos}(\mtPole,\rhos),
\end{equation}
where $\rhos$ is defined as
\begin{equation}
  \rhos=\frac{2 m_0} {\sqrt{\sttj}}.
\end{equation}
The definition of the variable \rhos is similar to the variable $\rho
= 2 m_t/\sqrt{s}$ often used in inclusive top-quark pair production.
However, since we want to measure the top-quark mass, $m_t$ cannot be
used in the definition of \rhos.  Instead we use a scale $m_0$ of the
order of the top-quark mass. In the following we set $m_0=170$~GeV.
Note, that in principle an arbitrary renormalization scheme for the
top-quark mass can be chosen.  In this article we will restrict the
analysis to the pole mass scheme.  The conversion from the pole mass
scheme to other schemes is feasible albeit technically involved. We
leave this issue for future studies.
\begin{figure}
\includegraphics[width=\columnwidth]{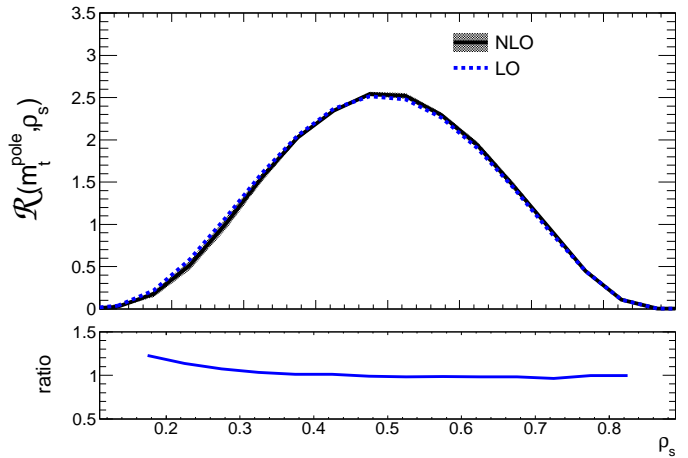}
\caption{Comparison between $\n3(\mtPole,\rhos)$
  calculated at LO and NLO accuracy for a $\mtPole=170$ GeV and using
  the CT09MC1 and CTEQ6.6 PDF sets, respectively.
  Below we show the ratio NLO/LO.\label{fig:LOversusNLO}}
\end{figure}
In \Fig{fig:LOversusNLO} we show LO and NLO predictions for the
distribution \n3 as defined in \Eq{eq:n3Definition}. LO and NLO
distributions are normalized to the respective LO and NLO cross
sections. Given that the area under both curves is equal to one the
curves need to cross which happens at $\rhos\approx0.45$. We observe
that in the region close to the threshold which corresponds to
$\rhos\approx 1$ the LO and NLO predictions agree reasonable
well. In \Fig{fig:PDFdependence} we compare results for different PDF sets.
In black we show the default setup where we use the CTEQ6.6 PDF set. In green
the result for the MSTW2008NLO set is shown. As a consequence of the
normalization the PDF dependence essentially cancels and the curves
lie on top of each other. \Fig{fig:PDFdependence} also shows the
uncertainty due to scale variations (black band). Evidently the scale uncertainty
is further reduced compared to the unnormalized quantities, since the
leading power in the strong coupling constant $\as$ cancels in the
ratio. One may argue that in such a situation the scale variation does
not provide a sensible method to estimate the effect of uncalculated
higher order terms. As a cross check we have compared the naive way to
calculate \n3 where we just divide the differential distribution by
\sigmattj with the expanded version where we perform a strict
expansion of \n3 in \as. We find that both methods lead
to roughly the same estimate for the uncertainty. 
We note that using the running top-quark mass instead of the
pole mass could in principle lead to a further reduction of the
scale uncertainty.
\begin{figure}
\includegraphics[width=\columnwidth]{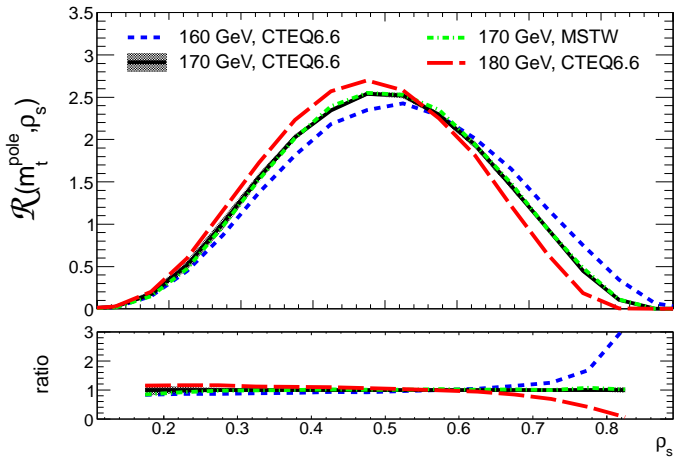}
\caption{ $\n3(\mtPole,\rhos)$ calculated at NLO accuracy for 
  different masses $\mtPole=160,\,170$ and $180$~GeV. For
  $\mtPole=170$~GeV the scale and PDF uncertainties evaluated as
  discussed in the text are shown. The ratio with respect to the
  result for $\mtPole=170$~GeV  is shown in the
  lower plot.\label{fig:n3MassDependence}}
\end{figure}
To investigate the sensitivity of the distribution \n3 to the
top-quark mass we have calculated \n3 for $\mtPole=160, 170, 180$ GeV.
The result is shown in \Fig{fig:n3MassDependence}. As before the three
curves need to cross since the area under each curve is normalized to
one.  The crossing happens slightly below $\rhos\approx 0.6$. At this point the
distribution is essentially insensitive to the top-quark mass. For
$\rhos\approx1$ we expect that the production of heavier quark
masses is suppressed compared to lighter masses. Indeed the
distribution for $\mtPole=180$ GeV is below the central curve while
the 160 GeV result lies above the result for 170 GeV. In the high
energy regime, that is for $\rhos\approx0$, we expect the opposite
to be true due to the normalization. For very large energies we observe
that the mass dependence is small as one would naively expect. From
\Fig{fig:n3MassDependence} we conclude that a significant mass
dependence can be observed for $0.4<\rhos<0.5$ and $0.7<\rhos$.  To
quantify the sensitivity we studied the quantity
\begin{eqnarray}
\label{eq:SDefinition}
{\lefteqn{
{\cal S}(\rhos) \,=\,}} \nonumber \\ 
& &\sum_{\Delta=\pm 5-10 \textnormal{\scriptsize ~GeV}}
\frac{|\n3(170~\GeV, \rhos)-\n3(170~\GeV+\Delta,\rhos)|}
{2 |\Delta|  \n3(170~\GeV, \rhos)}
\, .
\end{eqnarray}
\begin{figure}
\includegraphics[width=\columnwidth]{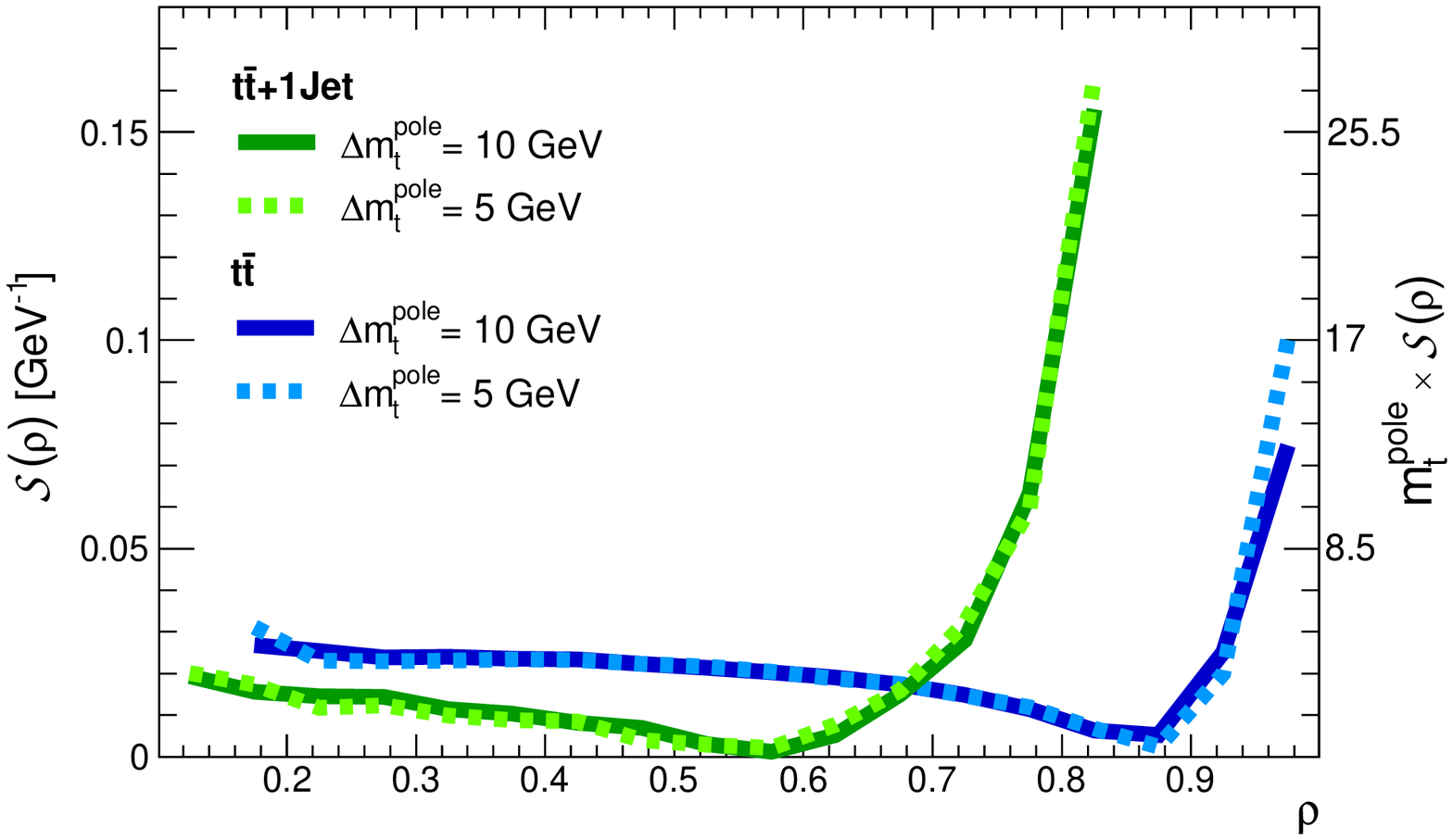}
\caption{
The sensitivity ${\cal S}(\rhos)$ of \n3 with respect to the 
top-quark mass  as defined in \Eq{eq:SDefinition}.
\label{fig:n3Sensitivity}}
\end{figure}
The result for ${\cal S}$ is shown in \Fig{fig:n3Sensitivity}. For
convenience the right $y$-axis shows $\mtPole \times {\cal S}$ which
is the proportionality factor relating the relative change in the
top-quark mass with the relative change in \n3:
\begin{equation}
  \left|{\Delta \n3 \over \n3}\right| \approx 
  \left(\mtPole {\cal S}\right)\,\times\, 
  \left|{\Delta \mtPole\over \mtPole}\right|.
\end{equation}
As can be seen in \Fig{fig:n3Sensitivity} values up to 25 are reached
for $\mtPole\times {\cal S}$ at $\rho\approx0.8$. With other words a
one per cent change of the mass translates into a 25 per cent change
of the observable \n3. The observable is thus five times more
sensitive than the inclusive cross section.  For comparison, in
\Fig{fig:n3Sensitivity}, we also show the sensitivity in case \n3 is
defined for the $t\bar t$ inclusive final state. (In the $t\bar t$ case we use
the definition $\rho=2m_0/\sqrt{s_{t\bar t}}$.) As one can see only
in the extreme threshold region---where reliable theoretical
predictions are challenging and also experimental uncertainties may
become large--- a similar sensitivity can be reached. Note that the
evaluation of the sensitivity relies on the assumption of a nearly
linear top-quark mass dependence. To cross check this assumption we
have used two different step sizes in \Eq{eq:SDefinition} (5 and 10
GeV). As can be seen from \Fig{fig:n3Sensitivity} the two results are
in perfect agreement. For a measurement not only the sensitivity is
important but also the expected theoretical and experimental
uncertainty. For example in the extreme threshold regime a good
sensitivity can be expected. However a reliable theoretical prediction
in that regime would require to go beyond fixed order perturbation
theory to resum threshold effects and soft gluon emission. To estimate
the impact of different uncertainties we show in
\Fig{fig:Uncertainties} the quantities
\begin{equation}
  \label{eq:rel3}
  \frac{\Delta\n3_\mu/\n3(170~\GeV,\rhos)}{{\cal S}(\rhos)} \mbox{ and } 
  \frac{\Delta\n3_{\textnormal{\scriptsize PDF}}/\n3(170~\GeV,\rhos)}{{\cal S}(\rhos)}
\end{equation}
where $\Delta\n3_\mu$ and $\Delta\n3_{\textnormal{\scriptsize PDF}}$ are the 
scale and PDF uncertainties of $\n3(172.5~\GeV,\rhos)$. We do not show
the region around $\rhos \approx 0.6$ because of the vanishing
sensitivity to the top-quark mass. \Fig{fig:Uncertainties} shows that
the main source of uncertainty comes from the scale variation while
the impact of the PDF uncertainties is much smaller.

From \Fig{fig:Uncertainties} we conclude that for $\rhos > 0.65$ the
dominant uncertainty is still below 1 GeV. The
low \rhos region, where \n3 loses its sensitivity to the top-quark mass,
could be used as control region for the experimental
reconstruction of \n3.

\begin{figure}
\includegraphics[width=\columnwidth]{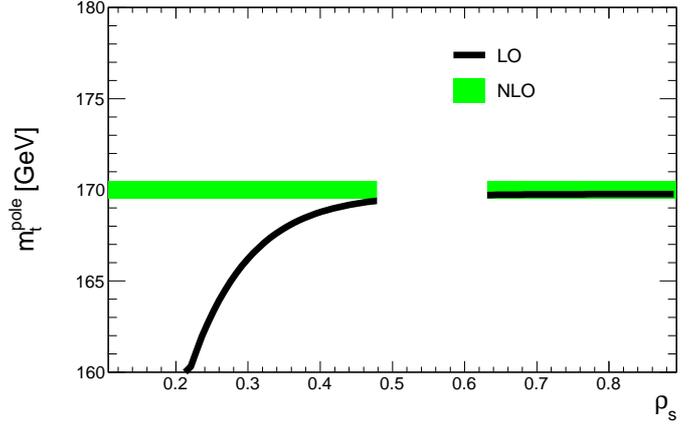}
\caption{The top-quark mass \mtPole  as obtained from a LO fit to 
  \n3 calculated in NLO accuracy (black line).  The input value of $\mtPole=170$
  GeV together with a variation of $\pm0.5$ GeV is shown as green
  band.\label{fig_LO}}
\end{figure}

\begin{figure}
\centering
\includegraphics[width=\columnwidth]{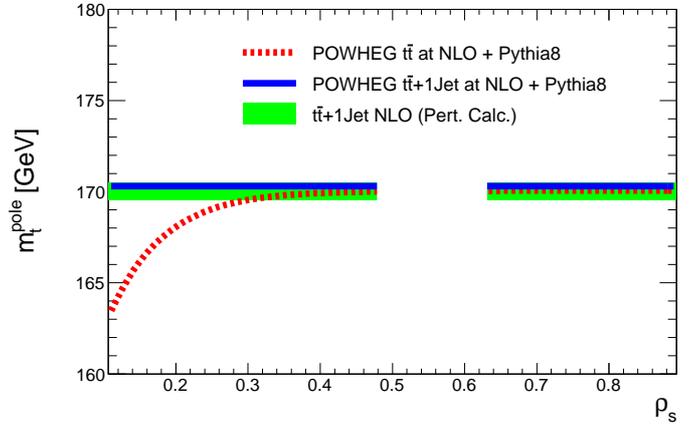}
\caption{Similar to \Fig{fig_LO} however $t\bar{t}$@NLO + POWHEG  
   and $t\bar{t}+\mbox{1-jet} + X$@NLO + POWHEG are used
  to fit the top-quark mass. \label{fig_POW}}
\end{figure}

To investigate further the impact of higher order corrections and the
effect of the parton shower we have compared the predictions for
$\n3(\mtPole,\rhos)$ using different approximations: LO,
$t\bar{t}$@NLO + POWHEG \cite{Frixione:2007nw}, \ttbaronejet@NLO
+ POWHEG \cite{Alioli:2011as}. In \Figs{fig_LO} and \ref{fig_POW} we
show the comparison as result of a toy experiment. We first calculate
\n3 in NLO accuracy at parton level. In a next step we compare with
the different approximations mentioned before and ask what top-quark
mass we would measure with a given approximation to explain the \n3
result calculated in NLO accuracy.  In \Fig{fig_LO} the comparison
with the LO calculation is shown. As input value we used $\mtPole=170
~\GeV$ in the NLO calculation. The thickness of the band denotes a
$\pm0.5$ GeV uncertainty. In the threshold region the LO result is
below the NLO curve (compare \Fig{fig:LOversusNLO}). As a consequence
we would fit a smaller top-quark mass to account for the deficit in
that region. The shift in the threshold region is about 2--3 GeV. In
the large energy regime the NLO result is below (see
\Fig{fig:LOversusNLO}) the LO result. Since in this regime the mass
effect works in opposite direction the fit using LO predictions would
again yield a smaller top-quark mass. However the NLO corrections are
much larger compared to the threshold region. As a consequence the shift
in the top-quark mass is more significant.\\
In \Fig{fig_POW} the same analysis is shown for 
$t\bar{t}$@NLO + POWHEG  and
\ttbaronejet@NLO + POWHEG. Since the various
predictions agree very well with each other we find agreement of the
different approximations within ~500 MeV for the extracted top-quark mass.

From the above findings we conclude that \n3 shows a good sensitivity
to the top-quark mass while at the same time theory uncertainties
lead to small uncertainties in the reconstructed mass value. We are
thus lead to the conclusion that  from the theoretical perspective
\n3 provides an interesting alternative to existing methods for
top-quark mass measurements. 

\section{Experimental viability study}
\label{sec:3}
Evidently for a true measurement the nice theoretical properties of
\n3 discussed in the previous section are not sufficient to conclude
that a measurement with a certain precision can be achieved. In this
section additional properties of the \n3 distribution which may affect
the experimental analysis are investigated.  For a realistic study we
use only stable particles in the final state originating from typical
$t\bar{t}+\mbox{1-jet} + X$ events as produced in proton-proton
interactions at 7 TeV center-of-mass energy. In particular the
top-quark decay and hadronization are taken into account leading to
complicated event topologies similar to those reconstructed in real
experiments.  In the Monte Carlos studies we use only publicly
available tools and do not make any reference to a particular LHC
experiment.  Since detector effects are to a large extent generic we
believe however that this should be sufficient to assess the dominant
experimental uncertainties.  The sources of uncertainties included in
this study are those which usually have a high impact on similar
multi-jet topologies. They are related to: {\it top-quark
  identification, jet identification and jet-jet invariant mass
  reconstruction}. The results obtained are intended to qualitatively
illustrate the reach of the method. A real experimental determination
will necessarily need a more detailed and careful detector analysis.

In the SM the top quark decays almost exclusively into a $W$ boson and
a $b$ quark and consequently a top-quark pair will give rise to
two $b$ quarks and two $W$ bosons that (each one of them) will further
decay into two quarks or into a lepton ($\ell$) and a neutrino
($\nu$). This study only considers the so called semi-leptonic decay
channel which assumes that one of the two $W$ boson decays
leptonically whereas the remaining $W$ boson decays hadronically.  All
quarks produced will later evolve into the hadronization and decay
processes until stable particles are produced.  This semi-leptonic
channel has a very good balance between efficient identification and
event rate since roughly $8/27$ of all top-quark pair events decay
semi-leptonically (as usual we exclude the tau decay from the leptonic
decays). These final state configurations can be identified by the
presence of one high-$\pt$ lepton (in our case only electrons and
muons are considered), high missing transverse energy because of the
presence of the neutrino, two jets originating from the $b$ quarks and
at least two additional jets.

Unless it is stated differently the observable \n3 has been calculated
using event samples simulated with \ttbaronejet@NLO + POWHEG mat\-ched
with Pythia as the parton shower generator and the fragmentation
model. Jets are defined using the \FastJet package
\cite{Cacciari:2011ma} and the anti-kt algorithm
\cite{Cacciari:2008gp} with R=0.4 consistent with the perturbative
study in the previous section. Selected events are required to fulfill
the following conditions:

\begin{enumerate}
\item only one lepton ($\ell =e, \mu$) with $\pt >$ 25 GeV and  $|\eta| <$ 2.5;
\item missing transverse energy larger $30$ GeV to account for the
  presence of a neutrino;
\item a large transverse mass of the leptonic 
  system\footnote{$M^W_T=\sqrt{2 \pt^{\ell} \pt^\nu 
      (1-\cos(\phi^\ell - \phi^\nu))}$ where $\ell$ is the lepton
    and $\nu$ is the neutrino coming from the $W\rightarrow \ell
    \nu$.}
  $M^W_T> 35$ GeV produced by the W boson decay;
\item at least 3 jets with $|\eta| < 2.5$, one of them with $\pt > 50$
  GeV and the other two with $\pt > 25$ GeV;
\item two additional identified b-jets.
\end{enumerate}

The jet with the highest energy is associated with the additional jet emitted in the hard scattering 
and the other two are assumed to originate from the $W$ boson decaying
hadronically.

In addition to these conditions other constraints based on the event
topology are applied.  The invariant mass of the two non $b$-jets
are required to be compatible with the mass of the $W$ boson within 20\% and
the two reconstructed top-jet systems are required to have similar
masses within a range of precision also around 20\%. The missing
energy has to be compatible with a neutrino which, together with the
identified lepton can both be attributed to originate from the decay
of the $W$ boson.
These conditions guarantee a good reconstruction of the jet
energies though better tunings or more optimized methods are available
for this purpose. However more involved methods would imply using detailed
detector specific tools which are beyond the scope of this exercise.
It should be noticed that it is only at this stage when the correct
jet association is imposed to test the hypothesis that the event
originates from two top quarks. In fact the quantity $\rho_s$
considered depends on the invariant mass of all jets
and fermions and is therefore independent of misidentified jet
associations.

After this selection the \n3 distribution has been successfully
reconstructed and the impact of the following effects has been
studied: 
\begin{itemize}
\item the event generator and fragmentation model: POWHEG
  with Pythia versus MC@NLO \cite{Frixione:2002ik} with Herwig
  \cite{Corcella:2000bw,Corcella:2002jc},
\item backgrounds, mainly QCD, single top and $W$+jets,
\item the impact of a wrongly reconstructed jet energy,
\item the unfolding procedure to correct the \n3
  distribution to the perturbative partonic level,
\item the statistical error depending on the collected luminosity,
\item different modeling of color reconnection.
\end{itemize}
The comparison between POWHEG interfaced with Pythia and MC@NLO
interfaced with Herwig shows that the reconstructed \n3 distribution
is very stable independent of the choice of the parton shower
generator and the fragmentation model which are selected in the
simulation\footnote{Due to the unavailability of the
    $\ttbaronejet$ implementation in MC@NLO, this comparison was limited to
     a $t \bar t $ NLO event sample.}. The value of \mtPole which
is obtained using one or the other model is found to remain stable
within $0.20\pm0.20$ GeV for the high range of $\rho_s$.

The presence of a high $\pt$ lepton (cut 1) and missing transverse
energy (cut 2) reduces the QCD background. The $W$+jets and single top contamination
is reduced by applying the cuts 3,4 and 5. The final
background is thus expected to be well understood and can be kept low 
($5-10$\%)
allowing for an easy subtraction with very small impact on the final
uncertainties.

\begin{figure}
  \includegraphics[width=\columnwidth]{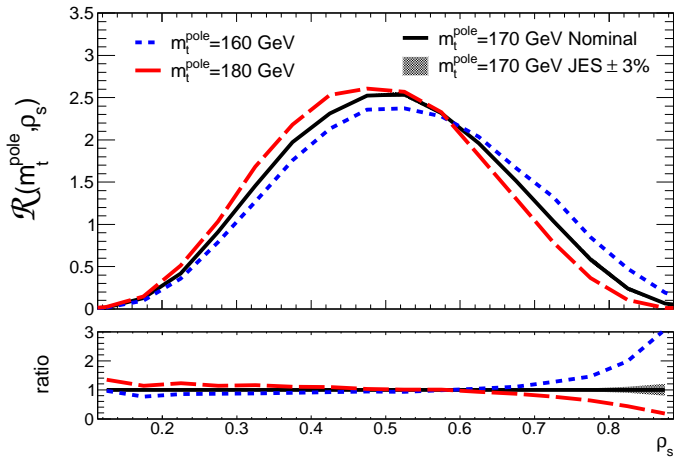}
  \caption{Impact of an uncertainty of $\pm$3\% in the jet energy scale.
    \label{fig:JESuncertainty}}
\end{figure}
The measured jet energy in the experiments can be distorted due to
different detection effects, such as, response of the calorimeters to
different particles, non-linearity responses of the detectors to the
particle energies, un-instrumented regions of the detector, energy
radiated outside the jet clustering algorithm, etc. The uncertainty
related to this kind of effects is often attributed to the jet energy
scale (JES). To estimate these effects the energy of the jets has been
changed by $3\%$ up and $3\%$ down before the construction of \n3.
The result is shown in \Fig{fig:JESuncertainty}. The black band
illustrates the uncertainty changing the JES scale by $\pm$3\%. As
reference we show also the result for \n3 using 160 and 180 GeV as
input for the top-quark mass.
Applying the topological constraints described previously,
the impact of this change to the top-quark mass measured using \n3
has been computed to be around $0.8-1.0$ GeV. A better
compromise between efficiency and resolution is possible though
it would imply tools which are detector specific.

\begin{figure}
  \includegraphics[width=\columnwidth]{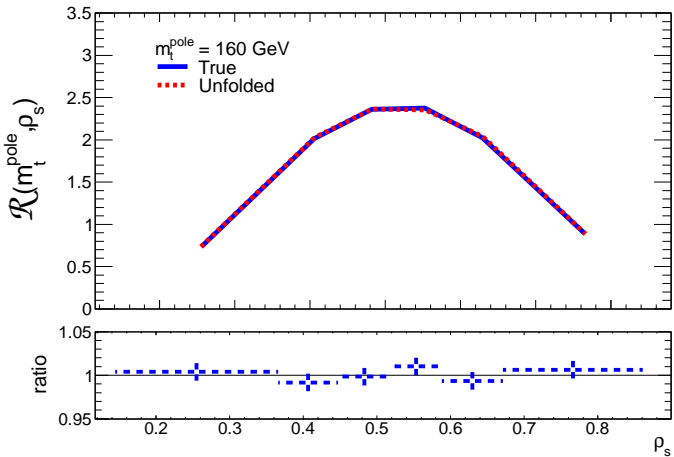}
  \caption{Mass independence of the unfolding procedure.
    \label{fig:unfolding}}
\end{figure}
It is important to realize that the observable proposed in this
article allows to unfold the data and reproduce the original \n3
distribution at parton level---without relying on a preset mass value
which would introduce what is often called the {\it Monte Carlo mass}.
A method based on the shape of the distribution has been developed for
this purpose. 
In \Fig{fig:unfolding} we illustrate the result of this procedure. The
unfolding procedure has been constructed using Monte Carlo simulations
for $\mtPole=170$~GeV. We have than applied the method to an event
sample generated for $\mtPole=160$~GeV. In \Fig{fig:unfolding} we compare
the result of the unfolding procedure with the Monte Carlo truth
information. As one can see the two curves lie on top of each
other. The uncertainty related to the unfolding is below one per cent.
The results thus show that a mass independent unfolding
is indeed possible in the region from $\mtPole=160$~GeV to
$\mtPole=180$~GeV. Due to the limited statistics used in this study
the previous statement has an associated uncertainty of $\sim$0.3 GeV.
The robustness of the method is thus reinforced by this result as it
shows that the correction procedure is not sensitive to the different
topological and phase-space configurations arising from the possible
different top-quark mass values. The influence of the minimum
gluon-jet $p_T$ has also been studied in the range 30 GeV $\leq p_T
\leq$ 60 GeV in this context. Again within the limited statistics we
find that the correction factors are stable at the level of 0.7\%.

To evaluate the statistical error we assume 5~fb$^{-1}$ collected
luminosity and a final efficiency of 1\% with respect to the original
\ttbaronejet cross section as shown in table
\ref{tab:LOandNLOxsections}.  The expected statistical error is around
1.4 GeV when integrating the high sensitivity region $\rho_{s}>0.65$
for this amount of collected luminosity.

\begin{figure}
  \includegraphics[width=\columnwidth]{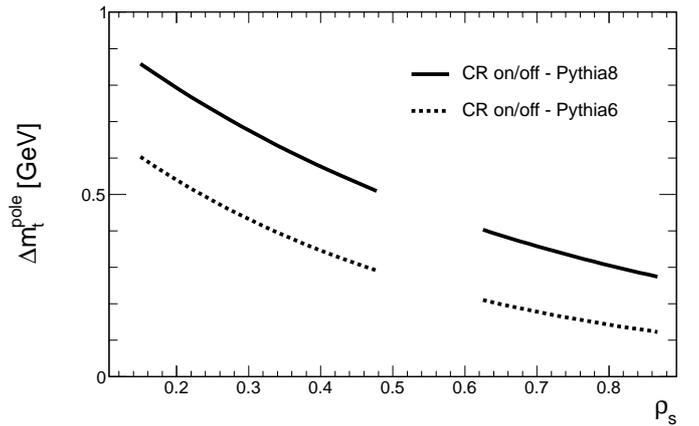}
  \caption{Impact of color reconnection on the top-quark mass
    determination using \n3. The solid line shows the effect using
    Pythia8 and switching color reconnection on and off. The dashed
    line shows the corresponding result using Pythia6.
    \label{fig:CRuncertainty}}
\end{figure}

As a last major uncertainty we have investigated the effect of color
reconnection. In the top-quark mass measurement via kinematical
reconstruction, the color reconnection directly affects the location
of the peak of the reconstructed distribution. As a consequence one
may expect effects of the order of 100 MeV. Indeed in
\Ref{Skands:2007zg} it is argued that an uncertainty of about 500~MeV
due to color reconnection can be expected. In the approach advocated
here we do not expect that color reconnection plays an important role
since the observable itself does not rely on a precise momentum
reconstruction. The momenta enter only through the jet algorithm which
determines whether an event passes the jet selection cuts. We expect
only a very weak effect of color reconnection on the jet algorithm.
Apart from this, color reconnection could in principle also affect the
determination of \sttj, however since \sttj is an inclusive quantity
we do not expect a major effect here. Furthermore an incorrect
determination of \sttj would only affect events at the bin boundaries
and is not unlikely that migrations at the left border and the right
border will compensate to a large extent. To study the impact of color
reconnection two different approaches have been investigated using the
Pythia versions Pythia6 and Pythia8 \cite{Sjostrand:2006za} which
consider two distinct color reconnection schemes.  In the Pythia8
color reconnection is assumed to happen before the top
quarks decay while in Pythia6, the color reconnection process is
performed after the top and $W$ decays. The main issue to consider is
the top lifetime and $W$ widths, and its relation with the time which
it takes the reconnection process to occur. To assess the
  impact of color reconnection on \n3 we used both Pythia6 and Pythia8 and
  compared the situation of using color reconnection in the respective
  default setup to the situation where color reconnection is
  completely switched off. The related uncertainty of the extracted
  top-quark mass is shown in \Fig{fig:CRuncertainty}.  Evidently the
  aforementioned procedure gives a rather extreme estimate of the
  uncertainty. We believe that the uncertainy given here is much more
  conservative compared to what has been done in \Ref{Skands:2007zg}
  where essentially only the impact of different tunes has been
  investigated. As a consequence the true uncertainty could be
  significantly smaller compared to what is shown in
  \Fig{fig:CRuncertainty}. In any case we conclude that even in the
  worst case the uncertainy is below 400 MeV using reasonable values
  for $\rhos$.

The above mentioned results prove the experimental
viability and the potential of the method. A real analysis using data
and detector specific tools is needed to understand the exact value of
the uncertainties affecting the determination of \mtPole but we can
estimate that a total error around of 1 GeV or lower is achievable.


\section{Conclusions}
\label{sec:conclusions}
In this article we propose a new method for top-quark mass
measurements at hadron colliders and in particular at the LHC. In 
detail the differential distribution of the \ttbaronejet cross section with respect to
$m_0/\sqrt{\sttj}$ is investigated. We
have shown that theoretical predictions for this quantity are well
under control and that the observable shows a good
sensitivity to the top-quark mass. Uncertainties related to
uncalculated higher order corrections or uncertainties in the parton
distribution functions are expected to affect the mass measurement by
less than 1 GeV. In a study of the experimental viability we have
addressed all major uncertainties without using a specific detector
set-up. Again we find that the impact on the top-quark mass
measurement is below 1 GeV. Since in the analysis presented here the
renormalization scheme of the top-quark mass is uniquely defined we
believe that the method nicely complements  existing approaches.

\section*{Acknowledgments}
We acknowledge discussions with R.~Nisius and S. ~Mart\'\i.  We also
thank T. Sjostrand for his comments and suggestions to study the color
reconnection effects and M.~Mangano for his comments on the
manuscript.  This work is partially supported by the Helmholtz
Alliance ``Physics at the Terascale'' HA-101, by the German Federal
Ministry for Education and Research (05H12KHE), by the Spanish
Ministry of Economy and Competitivity (FPA2012-39055-C02-01 and
AIC-D-2011-0688), by the German Research Foundation (DFG) through
 SFB-TR9 (B1),  and by the European Commission through contract
PITN-GA-2010-264564 ({\it LHCPhenoNet}).

\section{APPENDIX}

\begin{table}[h]
\begin{center}
\caption{Similar to \Tab{tab:LOandNLOxsections} but for a \pt cut of
  25 GeV in difference to 50 GeV used in \Tab{tab:LOandNLOxsections}.}
\label{tab:LOandNLOxsections2} 
\renewcommand{\arraystretch}{1.3}
\begin{tabular}{l|l|l}
\hline
 & \multicolumn{ 2}{|c}{$\sigma_{t\bar{t}+1-\textrm{\scriptsize jet}}$ (pb)}\\
 & \multicolumn{ 2}{|c}{$p_{T}(jet)>25 \,\textrm{GeV}, \,|\eta(jet)|<2.5$    }\\
\hline
\mtPole [GeV] & LO & NLO\\
\hline
$170$ & $91.827(8)^{+53}_{-31}$ & $77.7(1)^{-1}_{-9}$ \\ 
$172.5$ & $85.39(1)^{+50}_{-29}$ & $72.43(6)^{-2}_{-9}$ \\ 
\hline
\end{tabular}
\end{center}
\end{table}

\begin{table}[h]
\begin{center}
\caption{Similar to \Tab{tab:PSxsections} but for $\mtPole=172.5$ GeV.}
\label{tab:PSxsections2}
\renewcommand{\arraystretch}{1.3}
\begin{tabular}{|l|l|l|}
  \hline
  \multicolumn{ 3}{c}{ $\sigma_{t\bar{t}+1-\textrm{\scriptsize jet}}$ [pb]} \\
  \hline
  \multicolumn{ 1}{c|}{$t\bar{t}$} & without addtional PS & 
  \multicolumn{ 1}{c}{47.2(1)} \\ \cline{ 2- 3}
  \multicolumn{ 1}{c|}{NLO}  & PS by Pythia8 & \multicolumn{ 1}{c}{43,2(5)} \\ 
  \hline
  \multicolumn{ 1}{c|}{$t\bar{t}+\mbox{1-jet}$} & without addtional PS& 
  \multicolumn{ 1}{c}{44.9(5)} \\ \cline{ 2- 3}
  \multicolumn{ 1}{c|}{NLO} & PS by Pythia8 & \multicolumn{ 1}{c}{43,0(5)} \\ 
  \hline
\end{tabular}
\end{center}
\end{table}



\end{document}